\documentclass[twocolumn, prb]{revtex4}
\pdfoutput=1
\usepackage{graphicx}
\usepackage{epsfig}
\begin{document}
\title{Condensates induced by interband coupling in a double-well lattice }
\author{Qi Zhou,$^{1}$  J. V. Porto,$^{2}$ and S. Das Sarma$^{1}$}
\affiliation{$^{1}$Joint Quantum Institute and $^{2}$Condensed Matter Theory Center, Department
of Physics, University of Maryland, College Park, MD 20742\\
$^{2}$ National Institute of Standards and Technology,  
Gaithersburg, Maryland, 20899}
\date{\today}
\begin{abstract}

We predict novel inter-band physics for bosons in double-well optical lattices(OL). An intrinsic coupling between the $s$ and $p_x$ band due to interaction gives rise to larger Mott regions on the phase diagram at even fillings than the ones at odd fillings. On the other hand, the ground state can form various types of condensates,  including a mixture of single-particle condensates of both bands, a mixture of a single-particle condensate of one band and a pair-condensate of the other band, and a pair-condensate composed of one particle from one band and one hole from the other band. The predicted phenomena should be observable in current experiments on double-well OL. 
\end{abstract}
\maketitle\textit{}

\section{Introduction}
The last decade has seen many exciting developments in the studies of cold atoms in OL\cite{Blochr}. A number of many-body phases of bosons and fermions have been observed in experiments\cite{Bloch1, Spielman,  Nate, Bloch2, BlochM, Randy}, showing that cold atoms in OL provide an ideal platform for studying a number  
of theoretical models in condensed matter physics.  For example, bosons in OL have been used to study the single-band Bose-Hubbard model extensively in the past few years\cite{Bloch1,Spielman,  Nate}. 

Despite the fact that single-band models are usually good descriptions for atoms in OL, recent studies have found important influence from the higher bands\cite{Trey1, Bloch3}.  The leading-order effect of such inter-band couping is to renormalize the parameters of the single-band model\cite{Mueller, Wu, Hans, Cui}. For current experimental parameters,  the phase boundary is found to change slightly while the structure of the phase diagram remains the same\cite{Mueller}. It is therefore commonly assumed that inter-band coupling only leads to small quantitative changes in the quantum phase diagram of bosons in OL. 

A question then naturally arises, whether OL can be used to create many-body systems, which go beyond the description of the renormalized single-band models, i.e., whether situations exist where interaction induced inter-band coupling could give rise to qualitative new quantum phases not conceivable in the single-band picture?  It is easy to see that the chemical potential $\mu$ needs to be much smaller than the band gap $\Delta_g$  between the $s$ and $p$ band for invalidating the standard single $s$ band description.  In this Letter, we propose using a system in which $\Delta_g$ can be tuned to be comparable with or even smaller than $\mu$. A single-band description for this system is no longer valid, and novel quantum many-body phenomena due to inter-band coupling emerge.

The system we consider is bosons in a double-well lattice. In the literature, there has been studies on bosons in an excited band in double-well OL or standard OL\cite{cjw,cjw2}. Here we focus on a new aspect regarding the inter-band physics between the lowest two bands in a double-well lattice. From the experimental side, double-well lattices have recently been realized by several groups with different lattice geometries\cite{Trey2, Trey2p, Bloch4,dlm,dlm1,dlm2}. We focus on the simplest case, where a second laser  is added to a simple cubic optical lattice along the $x$ direction\cite{Bloch4}. The physics we discuss here can be easily generalized to other cases. By choosing the wave length of the second laser as half of the first one which creates the cubic lattice, a  superlattice forms with a double-well potential.  For appropriate choices of the relative positions of the two lattices, the potential is symmetric with respect to the center of each lattice site, as shown in Fig.(\ref{fig: fig1}a).

\section{Hamiltonian}

The potential of a double-well lattice in our case can be written as  $V(\vec{R})=\mathcal{V}_x(R_x)+\mathcal{V}_y(R_y)+\mathcal{V}_z(R_z)$, where $\mathcal{V}_x(R_x)=V_L\sin^2(\frac{\pi R_x}{d})-V_S\sin^2(\frac{2\pi R_x}{d})$, $\mathcal{V}_y(R_y)=V_y\sin^2(\frac{\pi R_y}{d})$, and $\mathcal{V}_z(R_z)=V_z\sin^2(\frac{\pi R_z}{d})$, $V_L$($V_i$) and $V_S$ are the amplitude of the long and short lattice along the $x$($i=y, z$) direction, $d$ is the lattice spacing. For this periodic potential,  the single particle energy $\epsilon_{\sigma}(\vec{ k})$ and the Wannier wave function $\psi_\sigma(\vec{R})$  can be exactly calculated, where $\sigma$ is the band index for the three dimensional lattice.  $\epsilon_{\sigma}(\vec{ k})$ can be written as the sum of the corresponding single particle energy $\epsilon_{i,n}(k_i)$,  where $i=x,y,z$ and $n$ is the band index,  for the one dimensional lattice $\mathcal{V}_i(R_i)$. Typical band structures for a standard optical lattice and a double-well lattice are shown in Fig.(\ref{fig: fig2}). It is clear that in a double-well lattice, $\epsilon_{x,0}(k_x)$  and $\epsilon_{x,1}(k_x)$ become very close to each other and are well separated from higher ones.  Thus we focus on the lowest two bands for the three dimensional lattice, the energy of which can be written as  $ \epsilon_s(\vec{k})=\epsilon_{x,0}(k_x)+\epsilon_{y,0}(k_y)+\epsilon_{z,0}(k_z)$, $\epsilon_{p_x}(\vec{k})=\epsilon_{x,1}(k_x)+\epsilon_{y,0}(k_y)+\epsilon_{z,0}(k_z)$.

\begin{figure}[htbp]
\begin{center}
\includegraphics[width=3.6 in]{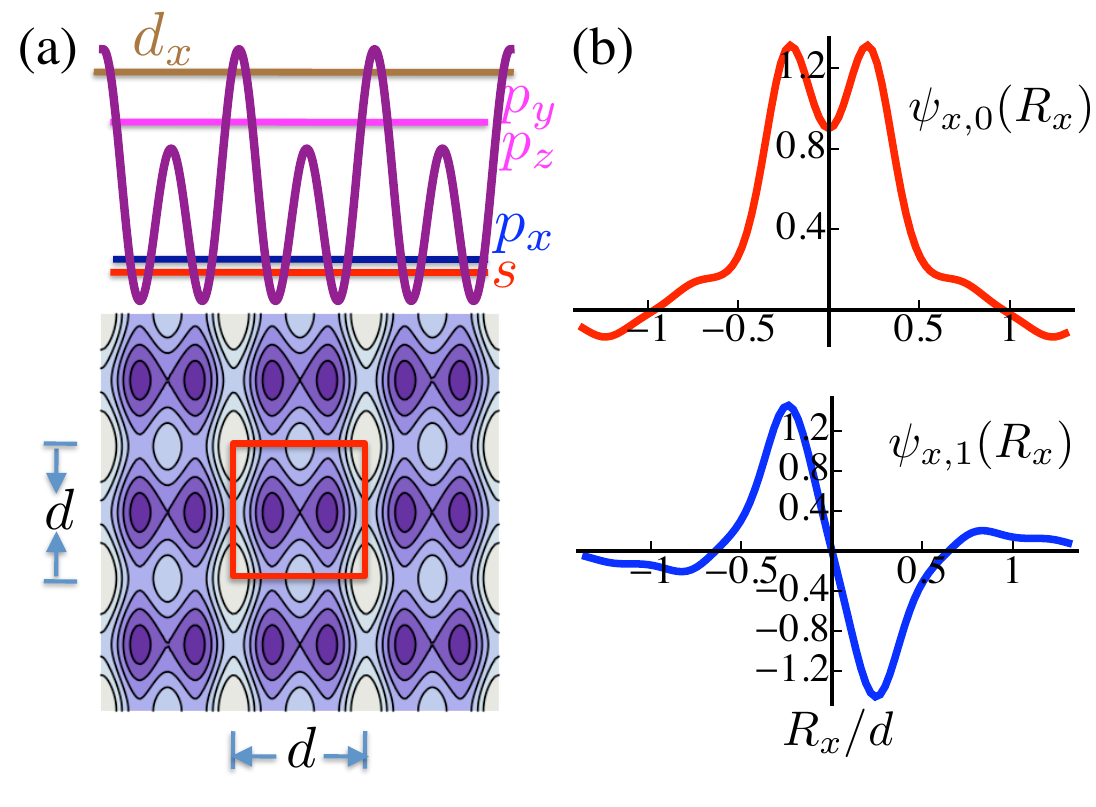}
\end{center}
\caption{(a) Schematic of a double-well lattice along the $x$ direction, as well as the lowest a few bands marked down in different colors. Contours with an unit cell marked in a red box on the $x-y$ plane are also shown. Dark region represents the potential minimum.(b) Wannier functions for the one dimensional lattice $\mathcal{V}_x(R_x)$, where  $V_S=11E_R, V_L=4E_R, V_y=V_z=13E_R$ along the $x$ direction.}\label{fig: fig1}
\end{figure}

The Wannier wave function for each band can be written as  $\psi_{s }(\vec{R})=\psi_{x,0}(R_x)\psi_{y,0}(R_y)\psi_{z,0}(R_z)$, $\psi_{p_x}(\vec{R})=\psi_{x,1}(R_x)\psi_{y,0}(R_y)\psi_{z,0}(R_z)$, where $\psi_{i,n}(R_i)$ is the Wannier wave function for the one dimensional lattice $\mathcal{V}_i(R_i)$ as shown in Fig(\ref{fig: fig1}b). Because the lattice potential $V(\vec{R})$ respects the inversion symmetry, i.e., $V(\vec{R})$ remains unchanged under the transformation $R_i\rightarrow -R_i$, $\psi_{\sigma}(\vec{R})$ has well defined parity.  For example, $\psi_{s }(R_x, R_y,R_z)= \psi_{s }(-R_x,R_y,R_z)$ and $\psi_{p }(R_x,R_y,R_z)= -\psi_{p }(-R_x,R_y,R_z)$. Both of them remain unchanged for $R_y\rightarrow -R_y$ or $R_z\rightarrow -R_z$.  This is similar to the parity of the usual $s$ and $p_x$ orbitals in a single harmonic oscillator. This is the reason that we denote the lowest two bands as the $s$ and $p_x$ bands for simplicity.

The many-body Hamiltonian can be written as $H=\sum_\sigma (H_\sigma-\mu_\sigma{\hat{N}}_{\sigma})+U_{sp}\sum_{{\bf m}}{\hat{n}}_{s{\bf m}}{\hat{n}}_{p{\bf m}}+H_c$,
where 
\begin{equation}
H_\sigma=\sum_{\sigma {\bf m}\vec{r}} t_{\sigma,\vec{r}}(\hat{b}^\dagger_{\sigma {\bf m}}\hat{b}_{\sigma {\bf m}+\vec{r}}+\text{c.c})+\frac{U_\sigma }{2}\sum_{\sigma {\bf m}}{\hat{n}}_{\sigma {\bf m}}(\hat{n}_{\sigma {\bf m}}-1)\label{Ha}
\end{equation}
is the usual Bose-Hubbard band model for the $s$ and $p_x$ band, and 
\begin{equation}
H_c=\sum_{{\bf m}}\left(W\hat{b}^\dagger_{p{\bf m}}\hat{b}^\dagger_{p{\bf m}}\hat{b}_{s{\bf m}}\hat{b}_{s{\bf m}}+\text{c.c}\right)\label{eHc}
\end{equation}
is the inter-band paring coupling term. In above two expressions, the band index $p_x$ has been replaced by $p$ for simplicity, $\vec{r}=d\hat{x},  d\hat{y}, d\hat{z}$ represents the unit vector along the $x,y, z$ directions, $\hat{b}^\dagger_{\sigma {\bf m}}$ creates a particle in the basis of Wannier wave functions at a lattice site ${\bf m}=d(m_x,m_y,m_z)$ that contains two wells, $\mu$ the chemical potential, $\mu_s=\mu$, $\mu_p=\mu-\Delta_g$, $\Delta_g$ is the energy difference between the middle of the $s$ and the $p_x$ band, which will be referred to as the band gap for simplicity.  $t_{\sigma,\vec{r}}=\sum_{\vec{k}} \epsilon_\sigma(\vec{k})e^{i\vec{k}\cdot \vec{r}}/N_s$ is the tunneling amplitude between the nearest neighbor sites, where $N_s$ is the total number of lattice sites. From exact numerical results, we found that tunneling amplitude beyond nearest neighbor sites is at least two orders of magnitude smaller, and thus can be ignored(See Appendix).  

\begin{figure}[tbp]
\begin{center}
\includegraphics[width=3.2 in]{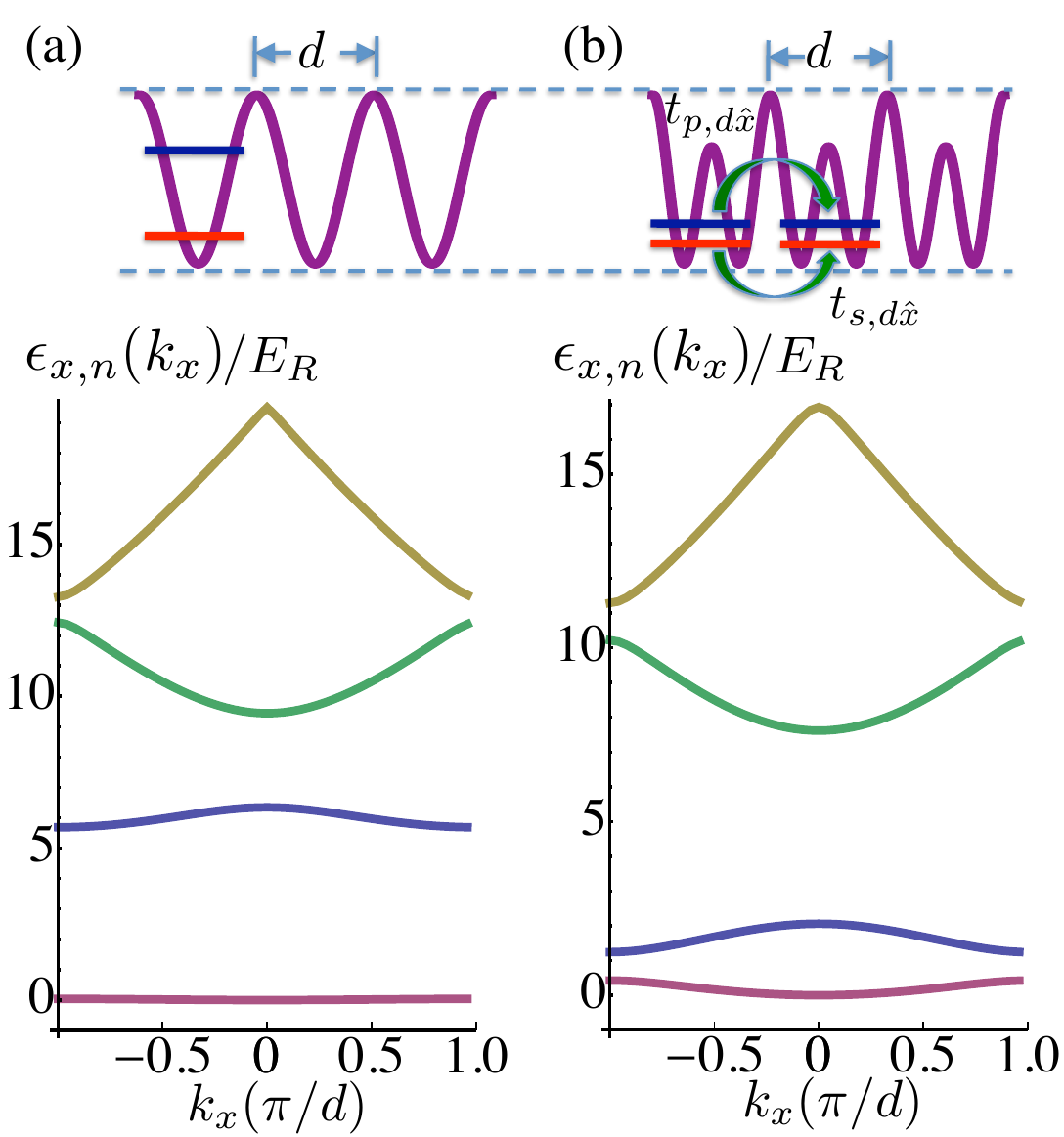}
\end{center}
\caption{(a) The energy spectrum for a standard lattice, where $V_L=13E_R, V_S=0, \Delta_g=6E_R$. (b) The energy spectrum for a  double-well lattice, where $V_L=4E_R, V_S=11E_R, \Delta_g=1.5E_R$ in 1D along the x direction. For comparison, the differences between the maximum and the minimum of these two lattices have been chosen to be the same. Red and blue horizontal lines in the upper panel represent the energy level of each Wanner orbitals. }\label{fig: fig2}
\end{figure}

 \begin{figure}[bp]
\begin{center}
\includegraphics[width=2.4 in]{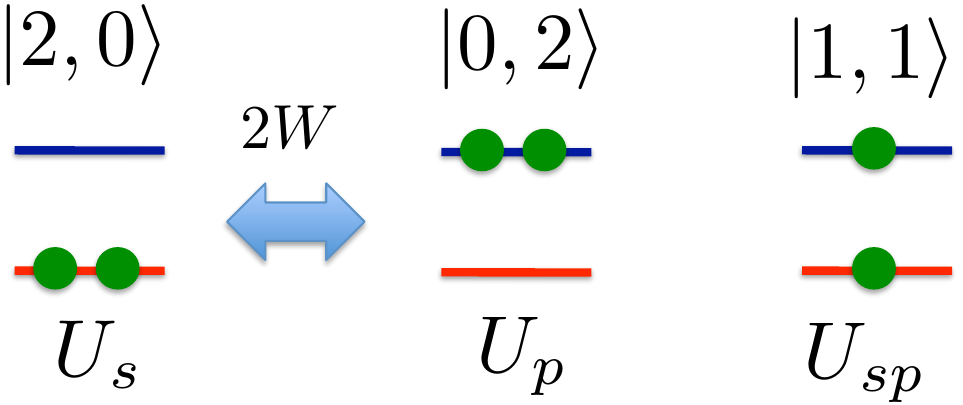}
\end{center}
\caption{Fock states and the matrix elements between them due to on-site interaction for two particles per site. }\label{fig: fig3}
\end{figure}
 
The interaction part in the Hamiltonian was obtained by expanding the field operator $\hat{\Psi}(\vec{R})$ in the general expression for the interaction of bosons $\mathcal{U}=\frac{2\pi\hbar^2a_s}{M}\int d^3R\hat{\Psi}(\vec{R})^\dagger\hat{\Psi}(\vec{R})^\dagger\hat{\Psi}(\vec{R})\hat{\Psi}(\vec{R})$ using $\hat{\Psi}(\vec{R})=\sum_{\sigma{\bf m}} \psi_{\sigma}(\vec{R}-{\bf m})\hat{b}_{\sigma{\bf m}}$. A straightforward calculation shows that there are three terms for onsite interaction. $\frac{U_\sigma }{2}\sum_{\sigma {\bf m}}{\hat{n}}_{\sigma {\bf m}}(\hat{n}_{\sigma {\bf m}}-1)$ is the intra-band repulsion for each band, where $U_\sigma=\frac{4\pi\hbar^2a_s}{M}\int d^3R|\psi_\sigma(\vec{R})|^4$. The second one is the inter-band repulsion, $U_{sp}=U_{sp}\sum_{{\bf m}}{\hat{n}}_{s{\bf m}}{\hat{n}}_{p{\bf m}}$, where $U_{sp}=\frac{8\pi\hbar^2a_s}{M}\int d^3R|\psi_s(\vec{R})|^2|\psi_p(\vec{R})|^2$. The last one as shown in Eq.(\ref{eHc}) describes the interacting process that scatters two atoms from the $s$ band to the $p_x$ band, where $W=\frac{2\pi\hbar^2a_s}{M}\int d^3R\psi^{*2}_s(\vec{R})\psi_p^2(\vec{R})$. Because of different parities of the $s$ and $p_x$ band, $\psi_s({R_x,R_y,Rz})=\psi_s(-{R_x},R_y,R_z)$, $\psi_p({R_x},R_y,R_z)=-\psi_p(-{R_x},R_y,R_z)$, the Hamiltonian does not contain terms like $\hat{n}_\sigma\hat{b}^\dagger_\sigma \hat{b}_{\sigma'}$, where $\sigma\neq \sigma'$.  We have dropped off other terms involving inter-site interaction that are much smaller. The Fock states for two particles per site as well as the matrix elements between them due to on-site interaction are shown in Fig.(\ref{fig: fig3}).

\begin{figure}[tbp]
\begin{center}
\includegraphics[width=2.6 in]{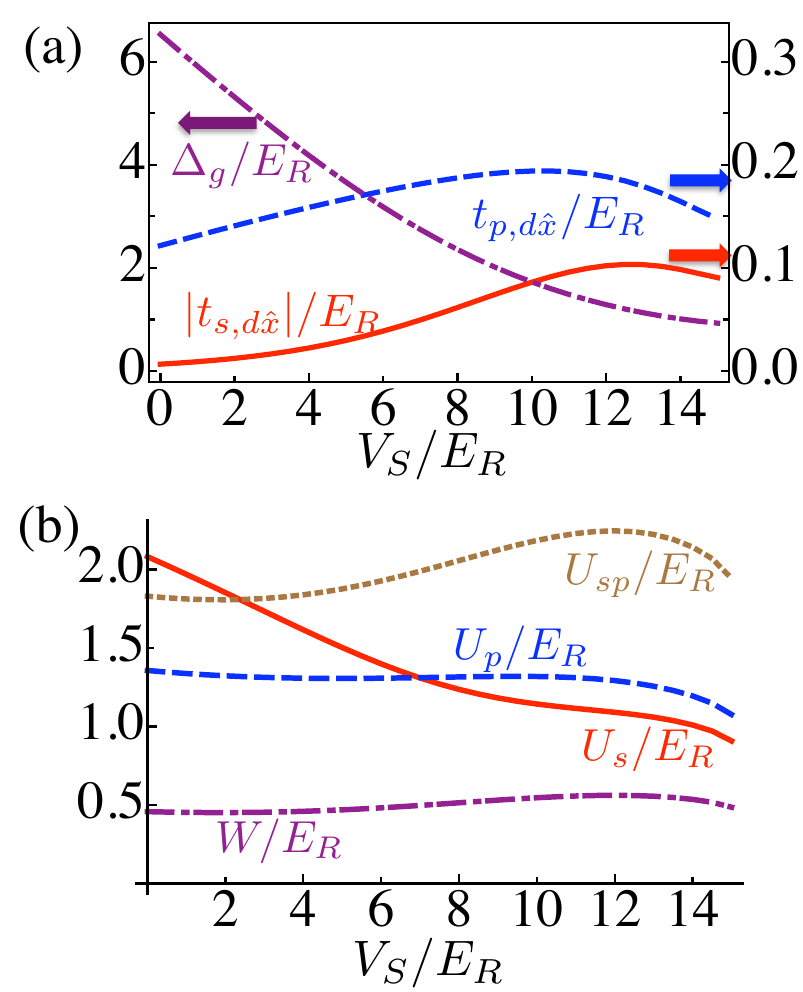}
\end{center}
\caption{ (a) $\Delta_g$(purple dash-dotted), tunneling along the $x$ direction of the $s$ (red solid) and $p_x$ band(blue dashed) as functions of $V_S$, where $V_S+V_L=15E_R$. (b) interaction $U_s$(red solid), $U_p$(blue dashed), $U_{sp}$ (brown dotted) and $W$(purple dash-dotted), where $a_s/d=0.05$ and $V_y=V_z=25E_R$.  }\label{fig: fig4}
\end{figure}

The Hamiltonian in Eq.(\ref{Ha}) has earlier been considered for standard OL with a fixed band gap at a constant lattice depth in a context different from our interest \cite{HZ,HZ2, ND}. Dependences of the parameters of the Hamiltonian in Eq.(\ref{Ha}) on $V_L/E_R$ and $V_S/E_R$ for a typical double-well lattice are shown in Fig.(\ref{fig: fig4}), where $E_R=h^2/8Md^2$ is the recoil energy. It is clear that as $V_S$ increases, $\Delta_g$ decreases quickly. 

\section{ Mean-field phase diagram} 

To solve this problem, the Gutzwiller mean field approach is used. The trial wave function is written as $|\psi\rangle=\prod_{{\bf m}}\sum_{\alpha,\beta}c_{\alpha\beta}|\alpha,\beta\rangle_{\bf m}$, where $|\alpha,\beta\rangle_{\bf m}$ is the on-site Fock state, $\alpha$ and $\beta$ are the particle numbers in the $s$ and $p_x$ orbital respectively. The order parameter for each band $\langle \hat{b}_{\sigma{\bf m}}\rangle=\phi_\sigma$ is determined self-consistently, and the phase boundary is obtained from $|\phi_s|=|\phi_p|=0$. 

\begin{figure}[tbp]
\begin{center}
\includegraphics[width=2.6 in]{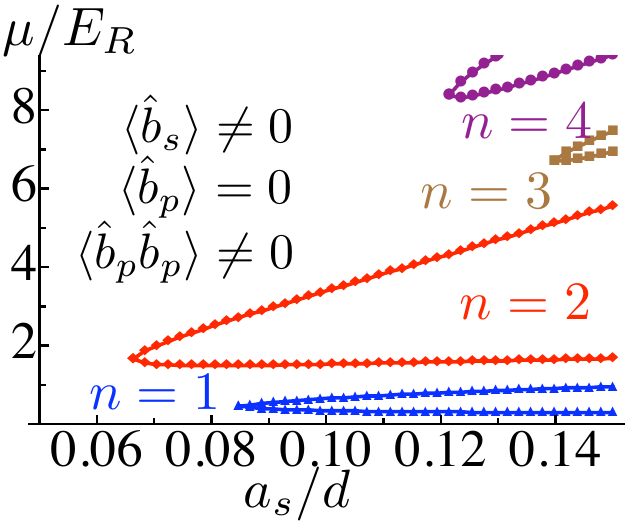}
\end{center}
\caption{Mean-field phase diagram, where $V_S=11E_R$, $V_L=4E_R$, $V_{y}=V_z=25E_R$. Phase boundary between a condensate phase and the Mott insulator is shown in different colors for different filling factors in the Mott insulator. }\label{fig: fig5}
\end{figure}

The phase diagram for the parameters $V_S=11E_R$, $V_L=4E_R$ and $V_{y}=V_z=25E_R$ is shown in Fig.(\ref{fig: fig5}). Compared with the usual results for the single-band Bose-Hubbard model,  it is clear that the structure of the phase diagram changes dramatically. For the single-band model, the area of the Mott region on the phase diagram monotonically shrinks with increasing filling factor. For our two-band model, there is an ``even-odd" effect, namely, the Mott region at an even filling $2n_0$  becomes larger than the one at an odd filling $2n_0-1$. 

This ``even-odd" effect can be first understood intuitively  using an extreme case, $V_L=0$ and $V_S\neq 0$. For this particular configuration, the Mott regions with odd fillings must vanish, since the lattice spacing has changed to $d/2$ and odd fillings in the original basis now correspond to non-integer fillings.  However, for any general case $V_L\neq 0$, the lattice spacing is still $d$. The above simple picture does not apply. To fully understand the ``odd-even" effect, one needs to consider on-site correlations induced by the inter-band coupling and the competition between the onsite energy and the tunneling energy.

The excitation gap of a Mott insulator can be written as $\Delta_n=E_{n+1}+E_{n-1}-2E_n$, where $E_n$ is the onsite energy of the Fock state with $n$ particles per site. In a single s-band model, $E^0_n=U_sn(n-1)/2-\mu n$. Consequently, $\Delta^0_n=U_s\sim a_s$ does not dependent on $n$, where the superscript $0$ represents the results for the single band model. On the other hand, the tunneling energy $-t_s\langle \hat{b}^\dagger_{s {\bf m}}\hat{b}_{s {\bf m+1}}\rangle\sim -t_s\sqrt{(n+1)n}$ increases with increasing $n$ due to bosonic enhancement.  As a result, the critical value of $a_s$, i.e., at the tip of the Mott region,  has to increase so as to overcome the increasing kinetic energy with increasing $n$ and form an insulator. This is the reason that Mott region monotonically shrinks on the phase diagram of the single band model with increasing $n$ or $\mu$. In our two-band model for double-well lattices, $H_c$ lowers the onsite energy $E_n$ by coupling the $s$ and $p_x$ band, $E_n=E_n^0+\delta E_n$, and modulate the nature of the excitation gap $\Delta_n=\Delta^0_n+\delta \Delta_n$. For example,  $\delta E_2=-4W^2/(2\Delta_g+U_p-U_s)+O(W^4) $ and $\delta E_3=-12W^2/(2\Delta_g+U_p+2U_{sp}-3U_s)+O(W^4)$ because of the formation of the pair $c_{20}|2,0\rangle+c_{02}|0,2\rangle$ and $c_{30}|3,0\rangle+c_{12}|1,2\rangle$ respectively, where $c_{20}$($c_{30}$) and $c_{02}$($c_{12}$) have a phase difference of $\pi$. $\delta E_2$($\delta E_3$) can be viewed as an effective binding energy induced by the paring coupling $H_c$ with respect to $|2,0\rangle$($|3,0\rangle$). On the other hand, $\delta E_1=0$ as the state $|1,0\rangle$ or $|0,1\rangle$ does not couple to any other Fock states. One can easily verify that $\delta \Delta_1<\delta \Delta_2$  if $U_{sp}>U_s$, where $\delta \Delta_1=\delta E_2<0$, $\delta \Delta_2=\delta E_3-2\delta E_2$. Thus $\Delta_1=U_s+\delta \Delta_1$ is suppressed and meanwhile $\Delta_2=U_s+\delta \Delta_2$ is either enhanced or reduced less than $\Delta_1$, i.e., $\Delta_1<\Delta_2$. As seen from Fig.(\ref{fig: fig4} (b)), $U_{sp}>U_s$ is satisfied if $V_S>2.5E_R$. Moreover, $U_{sp}-U_s$ increases with increasing $V_S$, so does $\Delta_2-\Delta_1$. Though populating the $p_x$ band may increase the kinetic energy for $n=2$ compared with $n=1$, when $\Delta_1\ll\Delta_2$, this kinetic energy increase, as well as the bosonic enhancement, is no longer important, and the critical value of $a_s$ for $n=2$ Mott insulator becomes even smaller than then one for $n=1$. That is what we see on Fig.(\ref{fig: fig5} ). Similar analysis applies for $n=3, 4$, and so on.

This ``even-odd" effect can be directly observed from in-situ images of the density of trapped bosons in a double-well lattice. With increasing interaction, the Mott plateau with two atoms per site would first emerge when $V_S$ is sufficiently high. The above discussion can be generalized to arbitrary orbital numbers $n^o$ in each lattice site. When there are strong pairing couplings similar to $H_c$ between different orbitals, the Mott region on the phase diagram for $(n \bmod{n^o})=0$ will be larger than the one for $(n \bmod{n^o})=1,2,.., n^o-1$. Similar phenomenon also happens for a model containing two or three $p$ bands in standard OL\cite{Girvin}.

\begin{figure}[tbp]
\begin{center}
\includegraphics[width=2.6in]{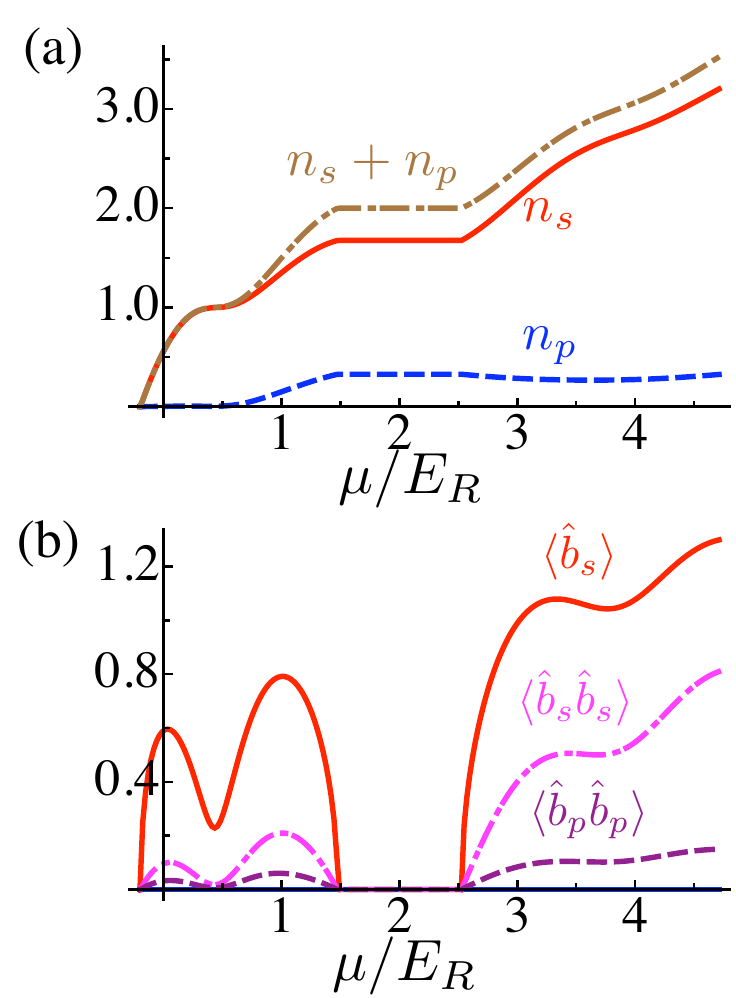}
\end{center}
\caption{(a) $n_s+n_p$(brown dash-dotted), $n_s$(red solid) and $n_p$(blue dashed) versus $\mu$ at ${a_s}/{d}=0.081$ for Fig.(\ref{fig: fig5}). (b) $\langle \hat{b}_s\rangle$(red solid), $\langle \hat{b}_s\hat{b}_s\rangle$(pink dash-dotted), $\langle\hat{b}_p\hat{b}_p\rangle$(brown dashed) and $\langle\hat{b}_p\rangle=0$ for same parameters as (a).}\label{fig: fig6}
\end{figure}

\section{ Condensate phases} 

Now we turn to the condensate phases.  Because of more than one orbital in each single site, there are multiple choices for the bosons to form condensate phases. (1) When $|\langle b_{s\bf m}\rangle| \neq 0$ and $|\langle b_{p\bf m}\rangle| \neq 0$, we call it  a ($\textit{C1}$) condensate, which is a coherent mixture of single-particle condensates of both bands. (2) When one of the single-particle condensates vanishes, we call it a ($\textit{C2}$) condensate, i.e., $|\langle b_{\sigma\bf m}\rangle|=0$ and $|\langle b_{\sigma'\bf m}\rangle| \neq 0$, we call it  a ($\textit{C2}$) condensate. (3). When $|\langle b_{s\bf m}\rangle|=|\langle b_{p\bf m}\rangle |=0$, the inter-band coupling induces a paired-condensate $|\langle  b^\dagger_{s\bf m} b_{p\bf m}\rangle|=|\langle  b^\dagger_{p\bf m} b_{s\bf m}\rangle|\neq 0$ under certain conditions.

\subsection{$(\textit{C1})$ condensate}

For a ($\textit{C1}$) condensate to emerge, the chemical potential $\mu$ needs to be larger than a critical value $\mu_c$. We write down the energy density for the excited band in the mean field approach as $E_p=-(\mu-\Delta_g+2\sum_{\vec{r}}t_{p\vec{r}})|\phi_p|^2+U_{sp}n_s|\phi_p|^2-2W|\phi_s|^2|\phi_p|^2+U_p|\phi_p|^4$. Based on Landau's criteria, $\mu_c=\Delta_g-2\sum_{\vec{r}}t_{p\vec{r}}+U_{sp}n_s-2W|\phi_s|^2$ and $|\phi_p|\neq 0$ for $\mu>\mu_c$.  In the parameter regime considered in Fig.(\ref{fig: fig4}), the very large $U_{sp}$ has largely increased the value of of $\mu_c$. Thus, the ground state forms a ($\textit{C2}$) condensate with $\langle \hat{b}_{s{\bf m}}\rangle\neq0$, $\langle \hat{b}_{p{\bf m}}\rangle=0$ and $\langle \hat{b}_{p{\bf m}} \hat{b}_{p{\bf m}}\rangle\neq0$. At lower lattice depths, i.e., $V_L=4E_R$ and $V_S=1E_R$, $U_{sp}$ is small enough and ($\textit{C1}$) condensates can emerge, i.e.,  for about four atoms per site\cite{QZ}.

\subsection{$(\textit{C2})$ condensate}

In double-well lattices, a ($\textit{C2}$)  condensate has a unique property: when $\langle \hat{b}_{\sigma{\bf m}}\rangle\neq0$,  $\langle \hat{b}_{\sigma'{\bf m}}\hat{b}_{\sigma'{\bf m}}\rangle$ must be finite(see Fig.(\ref{fig: fig6})). We can view the ($\textit{C2}$) condensate as a coherent mixture of a single-particle condensate of one band and a two-particle condensate in the other band.  A non-zero $\langle \hat{b}_{p{\bf m}} \hat{b}_{p{\bf m}}\rangle$ despite a vanishing $\langle \hat{b}_{p{\bf m}}\rangle$ results from the inter-band coupling $H_c$. Once $\langle \hat{b}_{s{\bf m}}\rangle\neq0$, the ground state includes Fock states $|\alpha,\beta\rangle_{\bf m}$ and $|\alpha\pm \nu,\beta\rangle_{\bf m}$, where $\nu$ is an integer.  On the other hand, the coupling term $H_c$ provides a matrix element between $|\alpha\pm2\nu,\beta\rangle_{\bf m}$ and $|\alpha,\beta\mp2\nu\rangle_{\bf m}$. The expectation value for $\hat{b}_{p{\bf m}}\hat{b}_{p{\bf m}}$ then must be finite, since the ground state includes both $|\alpha,\beta\rangle_{\bf m}$ and $|\alpha,\beta\pm2\nu\rangle_{\bf m}$, though it does not include Fock states $|\alpha,\beta\pm(2\nu+1)\rangle_{\bf m}$, i.e., $\langle \hat{b}_p\rangle=0$. This is very different from a single-band model, where $\langle \hat{b}\rangle$ and $\langle \hat{b}\hat{b}\rangle$ must vanish at the same time.  Here, $\langle \hat{b}_{\sigma'{\bf m}} \hat{b}_{\sigma'{\bf m}}\rangle$ can be induced by a finite $\langle \hat{b}_{\sigma{\bf m}}\rangle$ in a different band.  For standard OL, a ($\textit{C2}$) condensate also exists but the amplitude of $\langle \hat{b}_{p{\bf m}} \hat{b}_{p{\bf m}}\rangle$ is rather small because of the very large band gap. Our double-well lattices significantly reduce the band gap and make ($\textit{C2}$) condensate observable in experiments. 

A ($\textit{C2}$) condensate can be probed in Time-Of-Flight(TOF) experiments with a band mapping technique\cite{bm}, which maps the crystal momentum distributions in different bands, i.e., $\langle \hat n_{s,{\bf k}}\rangle$ and $\langle \hat n_{p,{\bf k}}\rangle$, separately to the first and second Brillouin Zone after expansion.  Since there is no single-particle condensate in the $p$ band, i.e., $\langle \hat{b}_{p{\bf m}}\rangle=0$, no singular feature is present in $\langle \hat n_{p,{\bf k}}\rangle$. On the other hand, a finite $\langle \hat{b}_{p{\bf m}}\hat{b}_{p{\bf m}}\rangle$ induces a novel structure in the momentum correlation function $I_p({\bf k}_1,{\bf k}_2)= \langle \hat{n}_{p,{\bf k}_1}\hat{n}_{p,{\bf k}_2}\rangle- \langle \hat{n}_{p,{\bf k}_1}\rangle\langle\hat{n}_{p,{\bf k}_2}\rangle$. A straightforward calculation shows, $I_p({\bf k}_1,{\bf k}_2)=I_1+I_2$, where $I_1=|\sum_{{\bf m}_1}\langle \hat{n}_{{\bf m}_1}\rangle e^{i({\bf k_1}-{\bf k_2})\cdot{\bf m}_1}|^2-\sum_{{\bf m}_1}\langle \hat{n}_{{\bf m}_1}\rangle^2$ and $I_2=|\sum_{{\bf m}_1}\langle \hat{b}^\dagger_{p,{\bf m}_1}\hat{b}^\dagger_{p,{\bf m}_1}\rangle e^{i({\bf k_1}+{\bf k_2})\cdot{\bf m}_1}|^2-\sum_{{\bf m}_1}|\langle \hat{b}^\dagger_{p,{\bf m}_1}\hat{b}^\dagger_{p,{\bf m}_1}\rangle|^2$. As a function of ${\bf k}_1$ and ${\bf k}_2$, $I_2$ has singular peaks at ${\bf k}_1+{\bf k}_2=q{\bf K}$, where $q$ is an integer, and ${\bf{K}}=\frac{2\pi}{d}(\hat{x},\hat{y},\hat{z})$ is the reciprocal lattice vector.  By measuring the amplitude of $I_p({\bf Q},-{\bf Q})$, where ${\bf Q}$ is an arbitary momentum in the second Brillouin Zone, the strength of $\langle \hat{b}^\dagger_{p,{\bf m}_1}\hat{b}^\dagger_{p,{\bf m}_1}\rangle$ can be probed. 

It is worthwhile to mention that, to observe $(\textit{C2})$ condensates discussed here, the temperature of the atoms needs to satisfy $T<t_s, W$. For a typical recoil energy $E_R\sim 100 nK$, it can be seen that $t_s$ and $W$ are of the order of tens $nK$. It is therefore quite promising to observe ($\textit{C2}$) condensates and the novel structure of the phase diagram in current experiments. 
 
\subsection{$(\textit{C3})$ condensate}

The Fock space in each single site can be divided to two subspaces, $\{S_1\}:$ $|n-2l,2l\rangle_{\bf m}$ and $\{S_2\}:$ $|n-2l-1,2l+1\rangle_{\bf m}$, where $l=0,1,2,...$, since $H_c$ only mixes states within each subspace. For $n$ particles per site, we can denote the ground state of each subspace as $|\Psi_{n1}\rangle_{\bf m}$ and $|\Psi_{n2}\rangle_{\bf m}$ and the ground state energy as $E_{n1}$ and $E_{n2}$.  Both $E_{n1}$ and $E_{n2}$ are dependent on $a_s$ and the lattice configuration. For the parameters in Fig.(\ref{fig: fig5}), $E_{n1}\ll E_{n2}$. On the other hand, under certain conditions, $E_{n1}\approx E_{n2}$. For example, for $V_S=1E_R$, $V_L=14E_R$, $V_y=V_z=80E_R$, which is essentially an one dimensional system, $E_{n1}\approx E_{n2}$ is satisfied for $n=3(5)$ around $a_s/d=0.18(0.123)$. Under this situation, by taking into account higher order correlations between nearest neighboring sites, an exotic ($\textit{C3}$) condensate arises.

 \begin{figure}[tbp]
\begin{center}
\includegraphics[width=2.8 in]{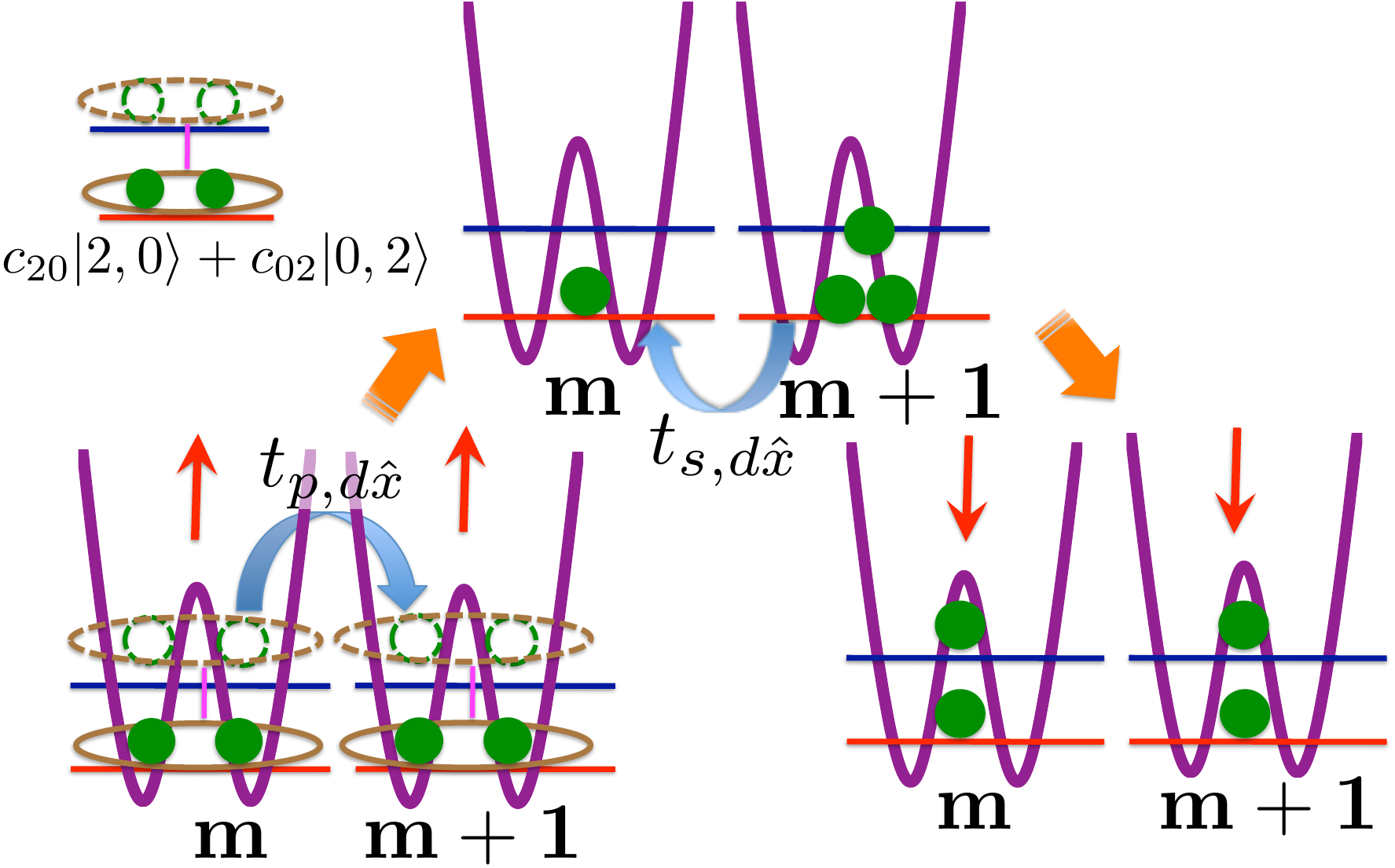}
\end{center}
\caption{ Schematic of a second order process that induces the correlations between the nearest neighbor sites. The bounded dimers represent the state $c_{20}|2,0\rangle+c_{02}|0,2\rangle$. The intermediate state with one and three particles in the nearest neighbor sites has much higher energy. The red vertical arrows represent the pseudospin in the spin model. }\label{fig: fig7}
\end{figure}

We apply a spin-${1}/{2}$ representation for the system, $|\uparrow\rangle_{\bf m}=|\Psi_{n1}\rangle_{\bf m}$ and $|\downarrow\rangle_{\bf m}=|\Psi_{n2}\rangle_{\bf m}$. The spin operators are defined as $S_{{\bf m}}^z=\frac{1}{2}(|\uparrow\rangle_{{\bf m}} {_{\bf m}}\langle\uparrow|-|\downarrow\rangle_{\bf m}{_{\bf m}}\langle{\downarrow}|)$, $S_{\bf m}^+=|\uparrow\rangle_{\bf m}{_{\bf m}}\langle\downarrow|$, $S_{\bf m}^-=|\downarrow\rangle_{\bf m}{_{\bf m}}\langle\uparrow|$, which satisfy the commutation relations, $[S_{\bf m}^+,S_{\bf n}^-]=2S^z_{\bf m}\delta_{{\bf m},{\bf n}}$,  $[S_{\bf m}^+,S_{\bf n}^z]=-S^+_{\bf m}\delta_{{\bf m},{\bf n}}$, $[S_{\bf m}^-,S_{\bf n}^z]=S^-_{\bf m}\delta_{{\bf m},{\bf n}}$. A spin-1/2 model can be obtained by applying second order perturbation theory(see Fig.(\ref{fig: fig7})), $H_{spin}=\sum_{{\bf m},\vec{r}}J_{\vec{r}}(S^+_{\bf m}S^-_{{\bf m}^+\vec{r}}+c.c)+\sum_{{\bf m},\vec{r}}\Delta_{\vec{r}}(S^+_{\bf m}S^+_{{\bf m} +\vec{r}}+c.c)+\sum_{{\bf m},\vec{r}}J^z_{\vec{r}}S^z_{\bf m}S^z_{{\bf m}+\vec{r}}-h\sum_{\bf m}S^z_{\bf m}$, where $J_{\vec{r}}, \Delta_{\vec{r}}\sim{t_{s,\vec{r}}t_{p,\vec{r}}}/(E^{ex}-E_{n1}-E_{n2})$,  $J^{z}_{\vec{r}}\sim {t_{s,\vec{r}}^2}/(E^{ex}-E_{n1}-E_{n2})+{t_{p,\vec{r}}^2}/(E^{ex}-E_{n1}-E_{n2})$, $h\approx E_{n2}-E_{n1}$ and $E^{ex}$ is the energy of an excited state with $n-1$ and $n+1$ particles on nearest neighbor sites. 

Unlike the usual $XXZ$ model obtained from a single-band Bose-Hubbard model, there is an additional term $(S^+_{\bf m}S^+_{{\bf m} +\vec{r}}+c.c)$ in our case, since the pseudospin is represented by Fock states with the same particle number and thus the total spin $S^z=\sum S^z_{\bf m}$ does not need to be conserved.  Because of this additional term, the spin model is no longer invariant under $U(1)$ gauge transformation, i.e., $S^+_{\bf m} \rightarrow S^+_{{\bf m}}e^{i \gamma}$,  where $\gamma $ is an arbitrary phase. Instead, only at discrete values of $\gamma$, i.e., $\gamma=\pm \pi$,  the spin model is invariant.  This indicates a $Z_2$ symmetry of this spin model.  

A mean field approach can be used to solve the above spin model. For example, when $J_{\vec{r}}, J^z_{\vec{r}}<0$, written down $\langle S^z_{\bf m}\rangle=\frac{1}{2}\cos\theta$, $\langle S^+_{\bf m}\rangle=\frac{1}{2}\sin\theta e^{i\varphi}$, $\langle S^-_{\bf m}\rangle=\frac{1}{2}\sin\theta e^{-i\varphi}$, the mean field energy of the spin model can be expressed as 
$\langle H_{spin}\rangle_M=\sin^2\theta( J_t+\Delta_t \cos{2\varphi})/2+(J^z_{t}\cos^2\theta)/4-(h\cos\theta)/2 $, where $J_{t}=\sum_{\vec{r}}{J_{\vec{r}}}$, $\Delta_t=\sum_{\vec{r}}{\Delta_{\vec{r}}}$ and $J^z_{t}=\sum_{\vec{r}}{J^z_{\vec{r}}}$. 
It can be easily verified that when $|J_t|+|\Delta_t|>|J_t^z|/2$ is satisfied, there is a critical value $h_c=|J_t|+|\Delta_t|-|J_t^z|/2$. If $|h|<h_c$, the value of $\theta$ to minimize $\langle H_{spin}\rangle_M$ is no longer zero or $\pi$, and there is a projection of the spin on the $x-y$ plane with a preference on the direction. i.e., $\varphi=\pm \pi/2$  for $\Delta_t>0$ or $\varphi=0, \pi$ for $\Delta_t<0$. Similar results can also be obtained for other signs of $J_{\vec{r}}, J^z_{\vec{r}}$.

We now go back to the boson representation $ |G\rangle =\prod_{\bf m}\left((\cos\frac{\theta}{2})^{\frac{1}{2}}|\Psi_{n1}\rangle_{\bf m}+(\sin\frac{\theta}{2})^{\frac{1}{2}}e^{i\varphi}|\Psi_{n2}\rangle_{\bf m}\right)$.
As $|\Psi_{n1}\rangle_{\bf m}=\sum_lc_{1l}|n-2l,2l\rangle_{\bf m}$ and $|\Psi_{n2}\rangle_{\bf m}=\sum_lc_{2l}|n-2l-1,2l+1\rangle_{\bf m}$, it can be seen that $\langle S^+_{\bf m}\rangle\neq 0$ implies $\langle \hat{b}^+_{s,\bf m} \hat{b}_{p, \bf m}\rangle=\langle \hat{b}^+_{p,\bf m} \hat{b}_{s, \bf m}\rangle\neq 0$. We refer to this state as a ($\textit{C3}$) condensate that emerges when $|h|<h_c$. It is a pair-condensate composed of one particle from one band and one hole from the other band. ($\textit{C3}$) condensates have also been found in a model describing two component bosons in a lattice, however, due to different underlying physics that belong to different universality class, i.e., breaking $U(1)$ other than $Z_2$ symmetry  as in our case\cite{Boris1}. We would like to point out that it is challenging to observe a ($\textit{C3}$) condensate in current cold atom experiments as it requires reaching a very low temperature, $T\sim t_\sigma t_{\sigma'}/U_\sigma$, which is of the order of $0.1-1nK$. 

Finally, the above spin model in 1D can be mapped to a fermion model via Jordan-Wigner transformation, $H_F=\sum_{\langle mn\rangle}J_{d\hat{x}}(\hat{c}^\dagger_{m}\hat{c}_{n}+c.c)+\sum_{{\langle mn\rangle}}\Delta_{mn}(\hat{c}^\dagger_{m}\hat{c}^\dagger_{n}+c.c)+4J^z_{d\hat{x}}\sum_{{\langle mn\rangle}}\hat{c}_m^\dagger\hat{c}_m\hat{c}_{n}^\dagger\hat{c}_{n}-(2h+4J^z_{d\hat{x}})\sum_{m}\hat{c}^\dagger_{m}\hat{c}_m$, where $\Delta_{mn}=-\Delta_{mn}=\Delta_{d\hat{x}}$. The fermion operators satisfy $\hat{c}_m=\left(\prod_{n<m}S^z_n\right) S^\dagger_m$, and $\hat{c}^\dagger_m=\left(\prod_{n<m}S^z_n\right) S^-_m$. Except for an unimportant nearest neighbor interaction term, this fermion model is identical to a model describing one dimensional p-wave superconductor\cite{Kitaev}. The degenerate ground states in the fermion model correspond to $\varphi=\frac{\pi}{2}/\frac{-\pi}{2}$ for $\Delta_t>0$ or $0/\pi$ for $\Delta_t<0$ in the boson model. 

\section{Conclusions}

We have shown that the band gap between the $s$ and one of the $p$ bands can be easily tuned in a double-well lattice. This tunable band gap gives rise to novel properties for bosons. Interaction induced inter-band coupling becomes crucial in this system and modifies  drastically the nature of the phase diagram, as well as  the possible condensate phases.  For the phase diagram, we found an ``even-odd" effect, namely Mott region with a filling factor $2n_0$ on the phase diagram becomes larger than the one with an odd filling $2n_0-1$. We also found three different types of condensate phases. Particularly, when strong interaction suppresses the fluctuations of the total density, inter-band coupling can nevertheless induce an paired-condensate that is composed by one particle from one band and one hole from another band.  Our results indicate the strong possibility of realizing novel condensate phases in double-well OL. Our studies can be generalized to the cases that more than one higher band is tuned to be very closed to the lowest band, where very rich physics is expected.\\

\section{Acknowledgements}
QZ thank K. Sun, H. Zhai, C. Wu and T.L. Ho for helpful discussions. This work is supported by JQI-NSF-PFC, ARO-DARPA-OLE, and ARO-MURI.

\section{Appendix}
 The kinetic energy in the momentum space can be written as $\mathcal{K}=\sum_{\sigma\vec{k}}\epsilon_{\sigma \vec{k}}\hat{a}^\dagger_{\sigma\vec{k}}\hat{a}_{\sigma\vec{k}}$,  where $a^\dagger_{\sigma\vec{k}}$($a_{\sigma\vec{k}}$) creates(annihilates) a particle in the basis of Bloch wave functions $\phi_{\sigma\vec{k}}(\vec{R})$.  Define $\hat{b}_{\sigma{\bf m}}^\dagger=\sum_{\vec{k}}\hat{a}^\dagger_{\sigma \vec{k}} e^{i\vec{k}\cdot{\bf m}}/\sqrt{N_s} $,  where $N_s$ is the total number of lattice sites, which creates a particle in the basis of Wannier functions $\psi_{\sigma{\bf m}}(\vec{R})$ at site ${\bf m}$, $
\mathcal{K}=\sum_{\sigma {\bf m}\vec{r}l} t_{\sigma,\vec{r},l}(\hat{b}^\dagger_{\sigma {\bf m}}\hat{b}_{\sigma {\bf m}+l\vec{r}}+\text{c.c})$,
where $l$ is an positive integer,  $\vec{r}=d\hat{x},  d\hat{y}, d\hat{z}$ represents the unit vector along the $x,y, z$ directions, and tunneling constants $t_{\sigma,\vec{r},l}=\sum_{\vec{ k}}\epsilon_{\sigma\vec{k}}e^{i l{\vec{k}\cdot \vec{r}}}/N_s$. Because the single particle energy is already diagonal with band indices, there is no crossing term involving tunneling between orbitals with different index $\sigma$. It can also be shown from a straightforward calculation that $t_{\sigma,\vec{r},l}$ corresponds to the overlap integral of the Wannier wave functions $\psi_{\sigma }(\vec{R}-{\bf m})$ and $\psi_{\sigma}(\vec{R}-{\bf m}-l\vec{r})$, i.e., $t_{\sigma,\vec{r},l} =\int d^3R\psi_{\sigma }(\vec{R}-{\bf m})^*(-\frac{\hbar^2\nabla^2}{2M}+V({\vec{R}}))\psi_{\sigma }(\vec{R}-({\bf m}+l{\vec{r}}))$.  From the exact band structure $\epsilon_{\sigma\vec{k}}$, we found that the tunneling amplitude between the nearest neighbor sites along any direction is always much larger than other ones. For example, for $V_L=4E_R$, $V_S=11E_R$, the one for the next nearest neighbor sites $t_{s,d\hat{x},2}$($t_{p,d\hat{x},2}$) are only $5\%$($2\%$) of $t_{s,d\hat{x},1}$($t_{p,d\hat{x},1}$) respectively. We thus keep only the tunneling between the nearest neighbor sites in the Hamiltonian. On the other hand, we have verified that introducing such a small correction from the tunneling between the next nearest neighbor sites does not change the qualitative feature of the phase diagram.  

There are also interaction terms for particles at different sites besides on-site ones. These terms, however,  depend on overlap integrals of Wannier wave functions at different sites, either in the same or different bands. Those integrals are always much weaker than the onsite interaction. For example, we have verified from exact numerical results that the largest nearest neighbor interaction is one or two orders of magnitude smaller than the onsite interaction for the parameters regime we considered. Thus we kept only onsite interaction in the Hamiltonian.

\end{document}